\documentclass[9pt,twocolumn,twoside]{opticajnl}
\journal{opticajournal} 

\setboolean{shortarticle}{true}

\usepackage{lineno}
\usepackage{upgreek}

\usepackage{soul}

\title{Photonic-crystal microresonator-based LiDAR engine}

\author[1]{Kenji Nishimoto}
\author[1]{Alexander E. Ulanov}
\author[1]{Thibault Wildi}
\author[1,2,*]{Tobias Herr}

\affil[1]{Deutsches Elektronen-Synchrotron DESY, Notkestr. 85,  22607 Hamburg, Germany}
\affil[2]{Department of Physics, Universität Hamburg UHH, Luruper Chaussee 149, 22761 Hamburg, Germany}

\affil[*]{tobias.herr@desy.de}

\begin{abstract}
Self-injection-locked (SIL) narrow-linewidth lasers based on high-Q microresonators are promising sources for frequency-modulated continuous-wave (FMCW) LiDAR, but the SIL mechanism as well as its key characteristics such as the frequency sweep range and the noise performance are often determined by uncontrolled backscattering in the resonator. Here, we investigate a tunable SIL laser based on a corrugated photonic-crystal (PhC) microresonator in which the feedback strength is set by design. Numerical and experimental results show that stronger SIL feedback expands the sweep range accessible through resonator modulation while also impacting the phase-noise and linewidth during sweeping, revealing a trade-off between frequency tunability and noise performance. Using CMOS-compatible microheater tuning (sub-1~V driving voltage), we demonstrate linearized up- and down-chirps with 224~THz/s over approximately 3~GHz and, in a proof-of-concept ranging experiment, measure a 10~m fiber length with a standard deviation below 3~mm. These results establish PhC microresonators with engineered SIL feedback as robust, compact, CMOS-compatible LiDAR engines.
\end{abstract}

\setboolean{displaycopyright}{false} 

\DeclareSymbolFont{CMletters}{OML}{cmm}{m}{it}
\DeclareSymbolFont{CMoperators}{OT1}{cmr}{m}{n}

\DeclareMathSymbol{\alpha}{\mathord}{CMletters}{"0B}
\DeclareMathSymbol{\zeta}{\mathord}{CMletters}{"10}
\DeclareMathSymbol{\eta}{\mathord}{CMletters}{"11}
\DeclareMathSymbol{\kappa}{\mathord}{CMletters}{"14}
\DeclareMathSymbol{\mu}{\mathord}{CMletters}{"16}
\DeclareMathSymbol{\xi}{\mathord}{CMletters}{"18}
\DeclareMathSymbol{\pi}{\mathord}{CMletters}{"19}
\DeclareMathSymbol{\tau}{\mathord}{CMletters}{"1C}
\DeclareMathSymbol{\psi}{\mathord}{CMletters}{"20}
\DeclareMathSymbol{\omega}{\mathord}{CMletters}{"21}

\DeclareMathSymbol{\Delta}{\mathalpha}{CMoperators}{"01}
\DeclareMathSymbol{\Lambda}{\mathalpha}{CMoperators}{"03}

\makeatletter
\def\@doi{}
\def\@journal@doi{}
\def\@article@doi{}
\makeatother

\date{}
\dates{}

\begin{document}

\maketitle

\noindent
\textbf{Introduction.} Narrow-linewidth, linearly tunable continuous-wave (CW) lasers are key to applications such as optical frequency domain reflectometry (OFDR)~\cite{tang2021hybrid}, spectroscopy~\cite{dmitriev2022hybrid}, and particularly frequency-modulated (FM) CW light detection and ranging (LiDAR)~\cite{lihachev2022low, snigirev2023ultrafast, lukashchuk2024photonic}. In FMCW LiDAR, target distance and velocity are retrieved by detecting the beat frequency between an up- and downward-sweeping CW laser frequency (up- and down-chirping) and its reflection from the target. By alternating the sweep directions, the range-dependent delay and velocity-dependent Doppler shift can be efficiently decoupled. Consequently, a wider frequency sweep range, lower laser noise, and higher sweeping linearity directly improve measurement precision and resolution~\cite{behroozpour2017lidar}, while faster sweeping speeds reduce the overall acquisition time. Semiconductor lasers with self-injection locking (SIL) via chip-integrated high quality (Q)-factor microresonators offer a compact, cost-effective, and energy-efficient platform to realize such sources~\cite{kondratiev2023recent, galiev2020optimization, jin2021hertz, li2021reaching}.

Recently, frequency-agile SIL lasers based on the hybrid integration of high-Q silicon nitride microresonators and piezoelectric materials have demonstrated rapid frequency tuning~\cite{lihachev2022low, lukashchuk2024photonic}. However, this approach relies on high-voltage piezoelectric actuators and involves complicated fabrication procedures for the co-integration of dissimilar materials. Furthermore, in standard integrated SIL lasers based on microring resonators, the accessible frequency sweep range and its noise performance critically depend on the coupling between forward- and backward-propagating waves~\cite{voloshin2021dynamics}. Because this coupling typically stems from random defect-based backscattering, achieving predictable performance and scalable fabrication is challenging. While previous studies have explored loop mirrors~\cite{shen2024reliable} or additional reflective sections~\cite{su2023self} to make the SIL feedback strength controllable, these methods often compromise the resonator Q-factor or increase the device footprint.

Here, we address these challenges by employing SIL based on synthetic reflection in a corrugated photonic-crystal (PhC) microresonator, combined with microheater modulation~\cite{wildi2024phase}. Unlike random defect scattering, the PhC structure introduces a deterministic, engineered reflection that provides precisely controllable SIL feedback strength without noticeably degrading the Q-factor~\cite{yu2021spontaneous, lucas2023tailoring, black2022optical, ulanov2024synthetic}. This flexibility allows us to deliberately tailor the feedback for SIL through chip design, reducing sensitivity to fabrication imperfections. We systematically investigate how this deliberately engineered coupling strength affects both the frequency sweep range and the phase-noise behavior. Our numerical and experimental results reveal that the corrugated PhC structure expands the accessible sweep range and turns the coupling strength into a reliable design parameter for balancing frequency tunability and linewidth reduction. With microheater-based tuning, we demonstrate a highly linear sweep rate of 224~THz/s over a range of nearly 3~GHz. As a proof-of-concept, we further validate this frequency-agile SIL source by performing FMCW-based length measurements of a 10~m-long fiber with a precision below 3~mm, demonstrating an efficient tuning mechanism that operates at significantly lower voltages than typical piezoelectric approaches.

\begin{figure}[!t]
\centering
\includegraphics[width=1.0\linewidth]{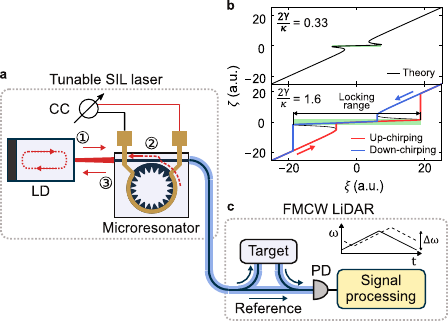}
\caption{(a)~Schematic of the tunable PhC microresonator-based SIL laser. The laser's emission frequency is swept by thermal tuning of the microresonator's resonance frequency via current control (CC) in an electric microheater. (b)~Examples of the SIL tuning curve calculated for $2\gamma/\kappa = 0.33$ and $1.6$. Red and blue curves show the system's tuning trajectories in up- and down-chirping tuning directions. The total accessible SIL range for laser tuning is highlighted in green. (c)~Basic configuration of the FMCW LiDAR system. Inset: Laser frequency $\omega$ received by the photodetector (PD) via the reference path (solid line) and the target path (dashed line).}
\label{fig:Figure_1}
\end{figure}

\noindent
\textbf{Theoretical description of tunable SIL.} Light emitted by the laser diode (LD) enters the microresonator as a forward-propagating wave (Fig.~\ref{fig:Figure_1}(a), \textcircled{\scriptsize 1}), where the PhC corrugation couples it to a backward-propagating wave at a rate $\gamma$, creating a synthetic back-reflection for SIL. The resonance frequency for which the PhC pattern creates the reflection signal and the reflection strength are defined by the spatial corrugation period, and the corrugation amplitude (Fig.~\ref{fig:Figure_1}(a), \textcircled{\scriptsize 2}); re-injection of the reflection into the LD cavity establishes SIL (Fig.~\ref{fig:Figure_1}(a), \textcircled{\scriptsize 3}). Via frequency pulling, SIL causes the laser to emit at the emission frequency $\omega$, which is different from the free-running LD emission frequency $\omega_{\rm d}$.
The relation between the detuning of $\omega_{\rm d}$ from the nearest microresonator resonance $\omega_{\rm m}$ normalized by $\kappa/2$ (where $\kappa$ is the total photon loss rate of the resonance), $\xi=\frac{2}{\kappa}\left(\omega_{\rm m}-\omega_{\rm d}\right)$, and the normalized effective detuning of $\omega$ from the same microresonator resonance, $\zeta=\frac{2}{\kappa}\left(\omega_{\rm m}-\omega\right)$, is given by~\cite{kondratiev2023recent, kondratiev2017self}

\begin{equation}
\xi=\zeta+K_{\rm 0}\frac{2\zeta{\rm cos}\psi\ +\left(1+(2\gamma/\kappa)^2-\zeta^2\right){\rm sin}\psi}{\left(1+(2\gamma/\kappa)^2-\zeta^2\right)^2+4\zeta^2}
\label{eq:Tuning_curve}
\end{equation}

where $\psi$ represents the phase of the feedback light returned to the LD, and $K_{0}$ is the combined coupling coefficient~\cite{galiev2020optimization},
\begin{equation}
K_{0}=\frac{8\eta\gamma}{\kappa^2}\sqrt{\Theta}\frac{T_{\rm o}^2 \sqrt{1+\alpha^2}}{\tau_{\rm d} R_{\rm e} R_{\rm o}^2}.
\label{eq:K0}
\end{equation}

Here, $\eta$ denotes the coupling coefficient between the microresonator and the bus waveguide ($\eta=0.5$ for critical coupling), $\Theta$ is the power mode coupling factor between the LD output and the waveguide, $T_{\rm o}$ and $R_{\rm o}$ are the transmission and reflection coefficients of the out-coupling mirror of the LD cavity, respectively, $R_{\rm e}$ is the reflection coefficient of the end mirror, $\tau_{\rm d}$ is the round-trip time of the LD cavity, $\alpha$ is the Henry factor. Figure~\ref{fig:Figure_1}(b) shows examples of SIL tuning curves for $2\gamma/\kappa=0.33$ and $1.6$, which can be calculated using Eq.~(\ref{eq:Tuning_curve}). Without PhC corrugation, the value of $\gamma$ in our resonator platform is randomly distributed, where values of around $2\gamma/\kappa =0.3\sim0.7$ are typical. 

Outside the SIL range, the lasing frequency is set by the free-running LD, i.e., $\omega=\omega_{\rm d}$ and $\xi=\zeta$. In contrast, on the stable locking branch (highlighted in green in Fig.~\ref{fig:Figure_1}(b)), $\omega$ is pulled towards $\omega_{\rm m}$, such that the laser remains close to resonance and $\zeta \approx 0$. 
The accessible SIL range is determined by the width of the stable locking branch, within which the slope of $\mathrm{d}\zeta/\mathrm{d} \xi$ is drastically reduced compared to the free-running case. This effect leads to an effective linewidth narrowing, and the linewidth-reduction factor is expressed as $\frac{1}{K^2}$, where $K=\frac{\mathrm{d}\xi}{\mathrm{d}\zeta}$ is the detuning sensitivity coefficient~\cite{galiev2020optimization}. A larger $\gamma$ increases the feedback and expands the accessible SIL tuning range. 

In our device, frequency tuning is realized by varying the electrical current applied to a microheater on the microresonator, which sweeps $\omega_{\rm m}$ mostly via the thermo-refractive effect. When the LD emission frequency $\omega_{\rm d}$ is not tuned, this sweep corresponds to a sweep of $\xi$ on the SIL tuning curve. For example, if $\omega_{\rm m}$ is swept towards higher frequency, the system follows the red line in Fig.~\ref{fig:Figure_1}(b). Since $\zeta \approx 0$ within the locking range, $\omega$ is also swept, tracking $\omega_{\rm m}$. Although synchronously sweeping $\omega_{\rm d}$ and $\omega_{\rm m}$ to constantly maintain $\xi \approx 0$ theoretically enables an even broader frequency tuning range, in this study, as in other tunable SIL lasers for FMCW LiDAR~\cite{lihachev2022low, lukashchuk2024photonic}, we prioritized operational simplicity and adopted only $\omega_{\rm m}$ tuning. 

For FMCW LiDAR, the frequency-swept laser signal can be split into a reference path and a target path (solid and dashed lines in the inset graph of Fig.~\ref{fig:Figure_1}(c), respectively), and the recombined signal can be detected by a photodetector (PD) for FMCW signal processing (see Supplementary Information (SI) Section 1 for details).

\begin{figure}[!t]
\centering
\includegraphics[width=\linewidth]{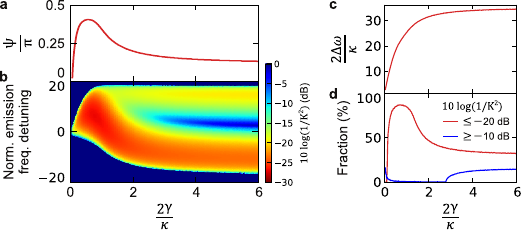}
\caption{(a)~Values for the feedback phase $\psi$ for which a maximal SIL range $\frac{2\Delta\omega}{\kappa}$ is reached. (b)~Linewidth reduction factor $\frac{1}{K^2}$ as a function of the emission frequency detuning and $\gamma$. (c)~Tuning range of the emission frequency as a function of $\gamma$. (d)~Fraction of the total SIL tuning range where the linewidth reduction factor satisfies either $10~\mathrm{log}\left(\frac{1}{K^2}\right)\leq -20~\mathrm{dB}$ or $10~\mathrm{log}\left(\frac{1}{K^2}\right)\geq-10~\mathrm{dB}$ as a function of $\gamma$.}
\label{fig:Figure_2}
\end{figure}

Since the frequency sweep range $\Delta\omega$ and laser linewidth of the SIL emission directly determine FMCW LiDAR performance, we first numerically investigate how $\Delta\omega$ and the linewidth-reduction factor $\frac{1}{K^2}$ vary with $\gamma$. For each value of $\gamma$, we extract the optimum $\psi$ that maximizes $\Delta\omega$ (Fig.~\ref{fig:Figure_2}(a)) and plot for this optimal value of $\psi$ the linewidth reduction factor (color-coded) as a function of normalized emission frequency detuning (Fig.~\ref{fig:Figure_2}(b)). Details of the numerical analysis are provided in the SI Section 2. Figures~\ref{fig:Figure_2}(c) and (d) summarize the resulting normalized sweep range $\frac{2\Delta\omega}{\kappa}$ (corresponding to the vertical width of the SIL region in Fig.~\ref{fig:Figure_2}(b)) and the fractions of the locking range with $10~\mathrm{log}\left(\frac{1}{K^2}\right)\leq -20~\mathrm{dB}$ and $10~\mathrm{log}\left(\frac{1}{K^2}\right)\geq-10~\mathrm{dB}$ for each $2\gamma/\kappa$. These results show that $\frac{2\Delta\omega}{\kappa}$ begins to saturate around $2\gamma/\kappa\approx2$, whereas the region with strong linewidth reduction decreases above $2\gamma/\kappa\approx0.7$. At the same time, a region with degraded linewidth reduction ($10~\mathrm{log}\left(\frac{1}{K^2}\right)\geq-10~\mathrm{dB}$) emerges around $2\gamma/\kappa\approx2.8$ and reaches about 15\% of the total $\Delta\omega$. These results show that although a larger $\gamma$ extends the SIL laser sweep range, it can also degrade the linewidth reduction and introduce a region with comparatively poor noise performance within the sweep range. This region arises because the resonance spectrum for a large $\gamma$ exhibits a pronounced split shape, which leads to a decrease in feedback strength within a specific detuning interval~\cite{ulanov2024synthetic}. Guided by these numerical results, we design and experimentally investigate SIL laser devices with different $\gamma$ to achieve a large $\Delta\omega$ while maintaining low-noise performance. We note that the optimal value of $\gamma$ for low-noise SIL lasers (without frequency sweeping) has been studied in Ref.~\cite{galiev2020optimization}.


\begin{figure}[!t]
\centering
\includegraphics[width=\linewidth]{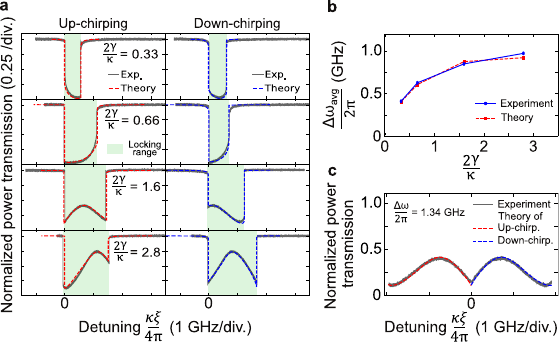}
\caption{(a)~Experimentally measured (solid) and numerically computed (dashed) normalized power transmission for up- and down-chirps. The green highlighted area indicates the SIL locking range (cf. Fig.~\ref{fig:Figure_1}(b)). (b)~Average sweep range $\Delta\omega_{\rm avg}$ as a function of $\gamma$. (c)~Experimentally measured (solid) and numerically computed (dashed) power transmission when the current sweep range is limited so that the system remains within the SIL range.}
\label{fig:Figure_3}
\end{figure}

\begin{figure}[!t]
\centering
\includegraphics[width=1.0\linewidth]{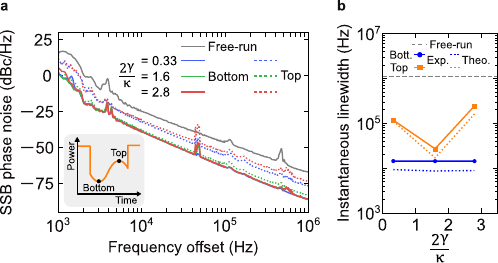}
\caption{(a)~Single-sideband (SSB) phase noise measured at bottom (solid) and top (dashed) operating points within the power transmission and in the free-running state. (b)~Comparison of the estimated instantaneous linewidth at the top and bottom operating points for each $2\gamma/\kappa$ (solid lines) and the instantaneous linewidth predicted from the free-running state linewidth and the linewidth reduction factor obtained by theoretical fitting (dashed lines).}
\label{fig:Figure_4}
\end{figure}

\noindent
\textbf{Experimental results.} For our experiments, we use microresonators characterized by a free spectral range (FSR) of 300~GHz and a Q-factor of $1.7\times10^6$. As conventional (non-PhC) reference cases, we select resonances with $2\gamma/\kappa=0.33$ and 0.66, corresponding to the range commonly obtained from random imperfection-based backscattering. We compare these with resonances for which a PhC corrugation intentionally increases $2\gamma/\kappa$ to 1.6 and 2.8. The values of $\gamma$ are estimated based on the observed resonance spectra (see SI Section 3). All resonances are in the wavelength range of $1557.8~\mathrm{nm}$ to $1560.2~\mathrm{nm}$, and are addressable by the same LD. The optical power coupled to the chip is approximately $500~\upmu\mathrm{W}$, well below 10\% of the parametric threshold power~\cite{herr2026FrequencyCombsCoherent} so that Kerr nonlinear effects~\cite{chermoshentsev2022dual} can be neglected. Across all experiments in Figs.~\ref{fig:Figure_3} and \ref{fig:Figure_4}, we keep the power input from the LD to the microresonator and all other terms in Eq.~(\ref{eq:K0}) constant, except for $\gamma$. 

Figure~\ref{fig:Figure_3}(a) shows both the experimentally measured (solid lines) and numerically calculated (dashed lines) normalized power transmission as a function of $\xi$ for different values of $2\gamma/\kappa$ while sweeping the resonance frequency $\omega_\mathrm{m}$ via the microheater modulation. The side-by-side results for up- and down-chirping show data acquired with the same sweep duration. The experimental results closely match the numerical model. We note that, for each value of $\gamma$, we fine-tune $\psi$ by adjusting the distance between the LD and the microresonator so as to maximize the average sweep range $\Delta\omega_{\rm avg}$ of the SIL laser over the up- and down-chirps. Details of the $\psi$ values selected in Fig.~\ref{fig:Figure_3}(a), the resonance spectra and the specific measurement method are provided in the SI Sections 2, 3, and 4, respectively.

Figure~\ref{fig:Figure_3}(b) plots the average up- and down-chirp sweep range $\Delta\omega_{\rm avg}/2\pi$ (blue solid curve), demonstrating that $\Delta\omega_{\rm avg}/2\pi$ increases as $2\gamma/\kappa$ increases, in agreement with the numerical model (red dashed curve). By limiting the heater-induced detuning sweep range, the chirp reversal occurs within the SIL locking range so that the SIL state is maintained throughout the entire detuning sweeps. 
Fig.~\ref{fig:Figure_3}(c) shows this for the case of $2\gamma/\kappa=1.6$. In this regime, as Fig.~\ref{fig:Figure_3}(c) shows, the power transmission curves for up- and down-chirping become symmetric, indicating that both chirps follow the same SIL tuning branch (highlighted in green in Fig.~\ref{fig:Figure_1}(b)). See SI Section 5 for details on $\psi$ during the SIL-maintained sweep.

Fig.~\ref{fig:Figure_4} shows the laser phase noise at the operating points corresponding to the strongest and weakest SIL feedback within the SIL range (bottom and top points of the transmission trace).
The phase noise is measured by detecting a beat note with a commercially available narrow-linewidth CW laser and utilizing digital IQ demodulation with an electrical spectrum analyzer (ESA)~\cite{kikuchi2012characterization}. The measurement results of the single-sideband (SSB) phase noise power spectral density (PSD) are shown in Fig.~\ref{fig:Figure_4}(a), where the solid and dashed lines represent the operating points at the bottom and top of the transmission trace, respectively, for $2\gamma/\kappa=0.33,~1.6,~2.8$, and the phase noise of the free-running LD is shown in gray. These measurement results demonstrate that the phase noise is in all cases reduced when operating at the bottom of the transmission trace, compared to operation at the top of the transmission trace.
In addition, Fig.~\ref{fig:Figure_4}(b) shows the corresponding instantaneous linewidths (see SI Section 6 for details), where the solid lines indicate the linewidths estimated from the experimental results, and the dashed lines indicate the values obtained by multiplying the free-running LD linewidth by the linewidth reduction factor obtained through the theoretical fitting in Fig.~\ref{fig:Figure_3}(a). In combination, the results in Figs.~\ref{fig:Figure_3}(b) and \ref{fig:Figure_4}(b) confirm the trade-off between a wide laser frequency sweep range and low-noise performance. In our case, the sample with $2\gamma/\kappa=1.6$ substantially expands the frequency sweep range and maintains strong suppression of linewidth degradation during the sweep. Thus, deterministic engineering of $2\gamma/\kappa$ using a PhC structure provides a practical approach to balance sweep range and linewidth in a controllable manner. 

\begin{figure}[!t]
\centering
\includegraphics[width=\linewidth]{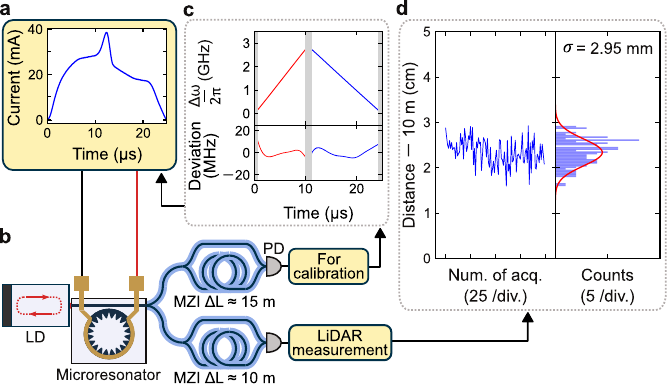}
\caption{(a)~Electrical heater current waveform for sweep linearization. (b)~Experimental setup; MZI: Mach-Zehnder interferometer. (c)~Frequency sweep (upper panel) and deviation from the ideal straight line (lower panel). (d)~Experimental results of FMCW measurements of a fiber delay line. Results of 100 consecutive distance measurements (left) and the corresponding histogram (right), with Gaussian fit (red line).}
\label{fig:Figure_5}
\end{figure}

As a proof-of-concept, we demonstrate an FMCW LiDAR ranging experiment, measuring the length of an optical-fiber delay line (Fig.~\ref{fig:Figure_5}(b)). In this experiment, we utilize the sample with $2\gamma/\kappa=1.6$ and a sweep duration for a complete up- and down-chirp of 25~$\upmu$s (40~kHz sweep rate). This value is chosen to balance sweep range and sweep rate, which is limited by the thermal response of the microresonator~\cite{nishimoto2022thermal}. To achieve a constant laser frequency-sweep rate, the electrical waveform modulating the microheater is not linear and deliberately pre-distorted with a peak-to-peak amplitude of \textasciitilde40~mA (equivalent to \textasciitilde$0.9~\mathrm{V_{pp}}$ based on a heater resistance of $\approx 23~\Omega$), as shown in Fig.~\ref{fig:Figure_5}(a). Additional details on the pre-distortion control waveform are provided in the SI Section 7. Further, to maximize the FMCW sweep range in this demonstration, we intentionally increase the power coupling $\Theta$ by reducing the spacing between the LD and the microresonator. This stronger power coupling enhances the locking term in Eq.~(\ref{eq:Tuning_curve}), thereby extending the detuning range over which the SIL state can be maintained. The power coupled to the resonator corresponds to 15\% of the parametric threshold power.
The linearized frequency sweep is shown in the upper panel of Fig.~\ref{fig:Figure_5}(c), and the lower panel displays the frequency deviation from an ideal linear chirp. The results of 100 consecutive FMCW measurements yield the fiber length with a standard deviation $\sigma=2.95~\mathrm{mm}$ for an overall length of approximately 10~m, as shown in Fig.~\ref{fig:Figure_5}(d). These results show that conventional pre-distortion-based linearization and ranging procedures can be directly applied to SIL sources based on a thermally-tuned PhC microresonator.

\noindent
\textbf{Conclusion.} 
In conclusion, we numerically and experimentally investigate a tunable self-injection-locked laser based on a CMOS-compatible PhC microresonator modulated by a microheater. We clarify how the engineered forward-backward coupling strength $\gamma$ affects the frequency sweep range $\Delta\omega$ and phase-noise behavior. We show that increasing $\gamma$ expands the laser frequency sweep range, but a trade-off exists between maximizing the sweep range and maintaining low-noise performance. In contrast to conventional non-PhC microresonators where SIL often relies on uncontrolled parasitic backscattering, PhC microresonators enable deliberate control over the value of $\gamma$, highlighting their potential as light sources for LiDAR. 
As a proof-of-concept, we demonstrate pre-distortion-based FMCW LiDAR leveraging a sample where a deliberately designed value of $\gamma$ balances sweep range and linewidth.
These results not only show that conventional FMCW LiDAR methods are applicable to PhC microresonator-based LiDAR engines, but also suggest that such sources could be combined with FMCW platforms using piezoelectric- or electro-optic-based modulation. 
Furthermore, synchronous modulation of the LD emission frequency and the resonance frequency may provide a route to further expanding the frequency sweep range.

\begin{backmatter}
\bmsection{Funding} This project has received funding from the EU's Horizon 2020 research and innovation program (grant agreements No 101137000 and No 101159229) and through the Helmholtz Young Investigators Group VH-NG-1404; the work was supported through the Maxwell computational resources operated at DESY.

\bmsection{Disclosures} The authors declare no conflicts of interest.

\bmsection{Data Availability Statement} Data underlying the results presented in this paper are not publicly available at this time but may be obtained from the authors upon reasonable request.

\bmsection{Supplemental document}
See Supplement 1 for supporting content.

\end{backmatter}

\bigskip

\bibliography{references}

\bibliographyfullrefs{references}


\end{document}